\documentclass[%
 reprint,
nofootinbib,
 amsmath,amssymb,
 prd,
]{revtex4-2}

\usepackage{aas_macros}
\usepackage{graphicx}
\usepackage{xcolor}
\usepackage{hyperref}
\usepackage{cleveref}

\newcommand{\R}{\mathcal{R}}
\newcommand{\E}{\mathcal{E}}
\newcommand{\Z}{\mathcal{Z}}
\newcommand{\K}{\mathcal{K}}
\newcommand{\Hu}{\mathcal{H}}
\newcommand{\Or}{\mathcal{O}}

\newcommand{\D}{\mathcal{D}}

\newcommand{\PP}{\mathcal{P}}

\begin{document}


\title{Primordial power spectra from $k$-inflation with curvature}

\author{Zakhar Shumaylov}
\email[]{zs334@cam.ac.uk}
\affiliation{Department of Applied Mathematics and Theoretical Physics, University of Cambridge,
Wilberforce Road, Cambridge, CB3 0WA, United Kingdom}
\author{Will Handley}%
\email[]{wh260@cam.ac.uk}
\affiliation{Astrophysics Group, Cavendish Laboratory, J.J.Thomson Avenue, Cambridge, CB3 0HE, United Kingdom}
\affiliation{Kavli Institute for Cosmology, Madingley Road, Cambridge, CB3 0HA, United Kingdom}

\date{\today}

\begin{abstract}
We investigate the primordial power spectra for general kinetic inflation models that support a period of kinetic dominance in the case of curved universes. We present derivations of the Mukhanov-Sasaki equations with a nonstandard scalar kinetic Lagrangian which manifests itself through the inflationary sound speed $c_s^2$. We extend the analytical approximations exploited in \citet{Contaldi_2003} and \citet{Thavanesan_2020} to general kinetic Lagrangians and show the effect of $k$-inflation on the primordial power spectra for models with curvature. In particular, the interplay between sound speed and curvature results in a natural low wave number cutoff for the power spectra in the case of closed universes. Using the analytical approximation, we further show that a change in the inflationary sound speed between different epochs in the early universe results in nondecaying oscillations in the resultant power spectra for the comoving curvature perturbation.

\end{abstract}

\maketitle

\section{Introduction}

An epoch of early accelerated expansion provided by cosmic inflation~\citep{Starobinskii1979, Guth1981, Linde1982} resolves several issues associated with naive hot big bang cosmologies, as well as predicts the spectrum of primordial curvature perturbations which we observe in the microwave sky today~\citep{planck_parameters, planck_inflation2018,2019PhRvD.100j3511H}.

A class of inflationary theory posits that there is ``just enough inflation,'' in contrast to the traditional chaotic or eternal models~\citep{kinetic_dominance, Hergt1, just_enough_inflation, just_enough_inflation2, BVS1, BVS2, Avis_2020}. Such theories are motivated by the presence of discrepancies at low multipoles~\citep{planck_isotropy}, the hints of present-day curvature which have been debated in the literature~\citep{2019arXiv190809139H, 2020NatAs...4..196D, 2020MNRAS.496L..91E} and their ability to possibly resolve~\citep{Ellis_2003, lasenbyclosed} some of the tensions between cosmological datasets which have emerged in the past few years~\cite{2019NatAs...3..891V,2021CQGra..38o3001D}. A Bayesian quantification of the degree of fine-tuning in such inflationary models can be found in detail in~\citep{Hergt2, Lukas_2020}. 

The work in this paper can be viewed as a generalisation and synthesis of the work summarised in \cite{Avis_2020,Contaldi_2003,Handley_2019, Thavanesan_2020, 2021PhRvD.104f3532G}:
\citet{Avis_2020} recently investigated phenomenologically the interaction between inflationary models with low speed of sound $c_s\ll1$ and spatial curvature. They demonstrated that in these models substantial large effects could be generated in proportion to $\Omega_K/c_s^2$, and hence such models could in theory be constrained by cosmic microwave background (CMB) data, or indeed even explain the discrepancies at low multipoles. However, in order to do this \citet{Avis_2020} identified that we must go beyond the linear order treatment considered there.
In order to explore this we generalise the derivation of the curved Mukhanov-Sasaki equation presented in \citet{Handley_2019} to include sound speeds that are not unity $c_s\ne1$ by considering the case of $k$-inflation~\cite{mukhanov1999}. The key result can be found in the generalised Mukhanov-Sasaki \cref{eq:R_main_conf,eq:MS} for a general kinetic Lagrangian $P(X,\phi)$ in the presence of curvature.

With the full $k$-inflation curved Mukhanov-Sasaki equation in hand, we may compute the primordial power spectrum numerically for any given Lagrangian. We showcase the generic effects of these models by deriving an analytical approximation first applied by \citet{Contaldi_2003}.  This approach is powerful as it is independent of the inflationary Lagrangian and hence isolates the generic properties of these models which specific choices of the Lagrangian modulate. \citet{Contaldi_2003} was generalised to the case of curvature by \citet{Thavanesan_2020} and to alternative initial conditions by \citet{2021PhRvD.104f3532G}, and this paper may be thought of as a further generalisation of that work.  As in \cite{Contaldi_2003,Thavanesan_2020}, we assume that the universe comes out of the big bang in a kinetically dominated regime and instantaneously switches to slow roll at some transition time $\eta_t$. In both epochs, the analytical solution can be found for the scale factor evolution, resulting in a closed form solution for the primordial power spectra.

The paper is structured as follows. In \Cref{sec:background} we derive the Mukhanov-Sasaki (MS) equation in the case of a curved universe and a general kinetic Lagrangian starting from the action. In \Cref{sec:ADM} we confirm the derived MS equation by expanding the action to second order starting from the ADM (Arnowitt-Deser-Misner) \cite{ADM} formalism. In \Cref{sec:analyt} we solve the MS equation using the analytical approximation. The power spectrum is calculated in \Cref{sec:discussion} from the approximation, the resulting solutions are shown for a variety of different parameters, and its relations to Planck 2018 data are discussed. Conclusions are presented in \Cref{sec:conclusion}.

\section{Background equations}\label{sec:background}
To ensure consistency in notation and introduce the base underlying equations, we begin by reviewing the derivation of the general dynamics of perturbations in single scalar field models. Similar derivations have been completed in e.g.~\citet{Garriga_1999,Baumann2012}.
\subsection{Zeroth order equations}
Working in the $+$\,$-$\,$-$\,$-$ metric, we begin with the general action 
\begin{equation}
    S=\int d^4x\sqrt{-g}\left[\frac12 R - P(X,\phi)\right],
\end{equation}
where 
\begin{equation}
    X = \frac12 g^{\mu\nu}\nabla_{\mu}\phi\nabla_{\nu}\phi.
\end{equation}
Extremising such an action produces the energy-momentum tensor 
\begin{equation}
    T^{\mu\nu}=-Pg^{\mu\nu} + P_{,X}\nabla^{\mu}\phi\nabla^{\nu}\phi,
\end{equation}
and equation of motion
\begin{equation}
    \label{eq:motion_1}
    \nabla_{\mu}(\sqrt{-g}P_{,X}\nabla^{\mu}\phi) + \partial_{\phi}\sqrt{-g}P=0,
\end{equation}
where subscripted $,X$ denotes the partial derivative with respect to $X$.
From the form of the energy-momentum tensor, we can identify the corresponding energy density and pressure, 
\begin{align}
    \label{eq:rho}\rho &=2XP_{,X} - P,\\
    \label{eq:p}p &= P. 
\end{align}
We will define the inflationary speed of sound as
\begin{equation}
    c_s^2\equiv\frac{p_{,X}}{\rho_{,X}} = \frac{P_{,X}}{P_{,X}+2XP_{,XX}} = 
    \frac{\rho+p}{2X\rho_{,X}}.
\end{equation}
The relevance of this definition will become apparent from the equation of motion for the curvature perturbations in the flat case, as it acts as the effective speed of sound.
We will consider scalar perturbations to the Friedmann-Lema\^{i}tre-Robertson-Walker (FLRW) metric in the Newtonian gauge:
\begin{equation}
ds^2=(1+2\Phi)dt^2-a^2(1-2\Psi)g_{ij}dx^idx^j,
\end{equation}
where 
\begin{equation}
g_{ij}dx^idx^j=\frac{dr^2}{1-Kr^2}+r^2\left(d\theta^2+\sin^2\theta d\phi^2\right).
\end{equation}
To zeroth order, the Einstein equations are equivalent to the following evolution equations:
\begin{align}
    H^2&=\frac13\rho-\frac{K}{a^2},\\
    \dot{H}&=-\frac{1}{2}(\rho+p)+\frac{K}{a^2}.
\end{align}
We can also expand to find time derivatives of the energy density and pressure:
\begin{align}
\dot{p}&=p_{,X}\dot{X}+p_{,\phi}\dot{\phi}\nonumber\\
\label{eq:p_evol}
&=-3c_s^2H(\rho+p)+\dot{\phi}(p_{,\phi}-c_s^2\rho_{,\phi}),\\
\label{eq:rho_evol}
\dot{\rho}&=-3H(\rho+p).
\end{align}

\subsection{First order equations}
Expanding the Einstein field equations perturbatively to first order gives us the following first order equations for the metric in the Newtonian gauge:

\begin{align}
\label{eq:r_0_1}
    \frac{d}{dt}\left(\frac{\delta\phi}{\dot{\phi}}\right)&=\left(1+\frac{2c_s^2(\nabla^2+3K)}{a^2(\rho+p)}\right)\Phi,\\
\label{eq:r_0_2}
    \frac{d}{dt}(a\Phi)&=\frac12a(\rho+p)\frac{\delta\phi}{\dot{\phi}},\\
    \Phi&=\Psi.
\end{align}
The 2 first order differential equations can be combined for a single second order equation of motion for the gauge invariant comoving curvature perturbation:
\begin{align}
    \R=\Psi+H\frac{\delta\phi}{\dot{\phi}}.
\end{align}
From \cref{eq:r_0_1,eq:r_0_2} above we can obtain
\begin{align}
\label{eq:R_main}
&(\D^2-K\E)\ddot{\R}+\left(\left(H+2\frac{\dot{z}}{z}\right)\D^2 - 3K\E H\right)\dot{\R} \\ &+ \frac{1}{a^2}\left(K(1+c_s^2\E-\frac{2}{H}\frac{\dot{z}}{z})\D^2 + K^2\E - c_s^2\D^4\right)\R = 0,\nonumber
\end{align}
where 
\begin{align}
    \E&=\frac{\rho+p}{2H^2 c_s^2},\\
    z&=\frac{a(\rho+p)^{\frac12}}{Hc_s},\\
    \D^2&=\nabla^2+3K.
\end{align}
It is more natural to analyse this equation of motion in conformal time $\eta$ defined through $dt=ad\eta$. Derivatives in conformal time will be denoted with $'$  and $\Hu\equiv\frac{a'}{a}$, in which case \cref{eq:R_main} becomes
\begin{align}
\label{eq:R_main_conf}
&(\D^2-K\E)\R''+\left(\frac{2z'}{z}\D^2 - 2K\E \Hu\right)\R'  \\ & +\left(K(1+c_s^2\E-\frac{2}{\Hu}\frac{z'}{z})\D^2 + K^2\E - c_s^2\D^4\right)\R = 0.\nonumber
\end{align}

\pagebreak
\section{Mukhanov-Sasaki equation}\label{sec:ADM}
Following~\cite{Handley2019,Thavanesan_2020} in this section we derive \cref{eq:R_main} starting from the action in the ADM metric, thus confirming our derivations. This form also allows us to directly canonically quantize the resulting perturbations. 

\subsection{ADM formalism}
We will begin with the line element in the ADM formalism:
\begin{equation}
ds^2=-N^2dt^2+g^{(3)}_{ij}(dx^i+N^idt)(dx^j+N^jdt)
\end{equation}
Note, that for the following section we will define the action as
\begin{equation}
    S=\int d^4x\sqrt{-g}\left[\frac12 R + P(X,\phi)\right],
\end{equation}
where 
\begin{equation}
    X = -\frac12 g^{\mu\nu}\nabla_{\mu}\phi\nabla_{\nu}\phi
\end{equation}
due to the signature of our line element being different. 
Such definitions have, for example, been adopted in~\cite{Cai2009}. This way, for a homogeneous scalar field $\phi(x^i,t)=\phi(t)$, the value of $X$ remains equal to $\frac12\dot{\phi}^2$ to zeroth order.

The action in the ADM metric becomes 
\begin{align}
\label{eq:action_ADM_1}
    S=\frac12\int d^4x\sqrt{-g}\bigg[NR^{(3)}+\frac{1}{N}(E_{ij}E^{ij}-E^2)+   2NP\bigg],
\end{align}
where 
\begin{equation}
    E_{ij}=\frac12(\dot{g}^{(3)}_{ij}-\nabla_iN_j-\nabla_jN_i),\,\, E=E_{\,\,i}^{i}.
\end{equation}
The corresponding Lagrangian constraint equations are:
\begin{align}
\label{eq:lag_constr_1}
    &R^{(3)}-N^{-2}(E_{ij}E^{ij}-E^2)=2\rho,\\
\label{eq:lag_constr_2}
    &\nabla_i(N^{-1}(E^i_j-\delta^i_jE))=0,
\end{align}
and expanding the metric perturbations to first order in the comoving gauge, we have $N=1+\alpha$, $N_i=\nabla_i\psi$, and $g^{(3)}_{ij}=a^2(1-2\R)c_{ij}$.
The Lagrangian constraint \cref{eq:lag_constr_1,eq:lag_constr_2} become
\begin{align}
    \alpha &=-\frac{\dot{\R}}{H}+\frac{K}{a^2}\frac{\psi}{H},\\
\frac{1}{a^2H}\D^2\R - \frac{1}{a^2}\D^2\psi &= -\frac{(\rho+p)}{2H^2c_s^2}\left(-\dot{\R}+\frac{K\psi}{a^2}\right).
\end{align}
The latter of which can be written as 
\begin{equation}
\label{eq:psi}
    \psi=\frac{\R}{H}-a^2\E(\D^2-K\E)^{-1}\left(\dot{\R}-\frac{K}{a^2}\frac{\R}{H}\right).
\end{equation}
Expanding the action in \cref{eq:action_ADM_1} gives (after integrating by parts)
\begin{align}
\label{eq:action_ADM_2}
&S=\frac12\int \sqrt{-c}a^3\frac{\rho+p}{H^2}\biggl[\frac{1}{c_s^2}\left(\dot{\R}-\frac{K}{a^2}\frac{\R}{H}\right)^2 - 
\\ &\frac{1}{c_s^2}\frac{K}{a^2}\left(\dot{\R}-\frac{K}{a^2}\frac{\R}{H}\right)\left(\psi-\frac{\R}{H}\right) - \frac{1}{a^2}\nabla_i\R\nabla^i\R + 3\frac{K}{a^2}\R^2 
\biggr].\nonumber
\end{align}
Plugging in from \cref{eq:psi} and integrating by parts once more gives
\begin{align}
\label{eq:R_action}
S&=\int d^4x\sqrt{-c}a^3\E\biggl[\frac{c_s^2}{a^2}\R\D^2\R+\\\nonumber&\left(\dot{\R}-\frac{K}{a^2}\frac{\R}{H}\right)\frac{\D^2}{\D^2-K\E}\left(\dot{\R}-\frac{K}{a^2}\frac{\R}{H}\right)\biggr].
\end{align}
There are a number of things to note about this: 
First, variation of the action in \cref{eq:R_action} with respect to $\R$ recovers the Mukhanov-Sasaki \cref{eq:R_main}.
Second, setting ${c_s^2=1}$ recovers the action shown in~\cite{Handley2019} and setting ${K=0}$ recovers the action shown in~\cite[Section 23.3]{Baumann2012}.

In analogy with~\cite{Handley2019}, the best we can do to get further in solving this is to define
\begin{equation}
    \Z = z\sqrt{\frac{\D^2}{\D^2-K\E}},\quad v=\Z\R.
\end{equation}
And rewriting the action in the new variable, we have
\begin{align}
S=\frac12\int d\eta d^3x\sqrt{-c}\biggl[v'^2+vc_s^2\D^2& v+\nonumber\\\left(\frac{\Z''}{\Z} + \frac{2K}{\Hu}\frac{\Z'}{\Z} - K\right)v^2\biggr]&.
\end{align}
Varying the action with respect to $v$ and Fourier decomposing the action, we recover the MS equation for the Mukhanov variable:
\begin{equation}
\label{eq:MS}
    v_k''+\left[c_s^2\K^2(k) -\frac{\Z''}{\Z} - \frac{2K}{\Hu}\frac{\Z'}{\Z} + K(1-3c_s^2)\right]v_k=0,
\end{equation}
where $\D^2$ is replaced with $-\K^2(k)+3K$ and 
\begin{align} 
    \K^2(k)=\begin{cases}
      k^2 \hfill k\in\mathbb{R}, k>0 & K=0,-1,\\
      k(k+2)\quad \hfill  k\in\mathbb{Z}, k>2 & K=+1.\\
    \end{cases}  
\end{align}

\pagebreak
\section{Analytical approximation}\label{sec:analyt}
We aim to solve \Cref{eq:MS} in order to connect the solutions to the observations from CMB. In most cases, it cannot be solved in generality, and one is normally forced to employ numerical schemes to solve it. This is a highly oscillatory differential equation, and due to that, normal numerical techniques suffer in both run-time and accuracy. There have been successful developments in solving such equations in the context of cosmological perturbations in, e.g.,~\cite{Agocs2020,Haddadin2021}. However, such techniques require one to specify the full Lagrangian and as a result make it difficult to notice general trends in resulting solutions. 
In order to simplify the calculations and make them potential agnostic, analytical approximations have been developed in~\cite{Contaldi_2003}, which were then extended for the case of curved universes in~\cite{Thavanesan_2020}. 

In this section we are going to extend the analytical approximation techniques to the case of a general scalar field Lagrangian with a "slowly" evolving speed of sound. As a result for the rest of the paper $c_s^2$ can be assumed to be practically constant. A rigorous approach would require one to define extra slow-roll parameters and more detailed derivations can be found in~\cite{Garriga_1999,Martin_2013,Huang_2013}.

The general approach is as follows:
\begin{enumerate}
    \item The evolution of the scalar field is assumed to be split into two distinct regimes: Kinetic Dominance~(KD) and Slow Roll~(SR). 
    \item During both epochs exact analytical solutions can be found for the evolution equations of the scalar factor. Solutions  to \cref{eq:MS} and their derivatives are then matched at some transition time $\eta_t$. Larger values of $\eta_t$ correspond to larger values of spatial curvature density at the onset of inflation.
    \item The power spectrum is calculated at the end of the SR epoch and then matched to Planck 2018 best-fit phenomenological parameters $(n_s,A_s)$. Note, that the parameters $(n_s,A_s)$ used throughout are best parameters for K$\Lambda$CDM, since for $k\to\infty$ we expect to recover the standard power spectrum for $c_s=$const, as  will be seen from \cref{eq:k_plus,eq:k-def}.
    \item The sound speed for curvature perturbations will be assumed to be slowly changing and can be taken to be constant but in the general case different between both epochs $c_{-}$ and $c_{+}$.
    \item KD can be loosely classified as $p/\rho=c_-^2$ and SR can be loosely classified as $\E\to0$. Reasons for and further implications of these assumptions will be discussed in more detail further below. 
\end{enumerate}
The overall scale factor evolution will be deduced from the following two equations:
\begin{align}
    \label{eq:Hus_evol_1}
    &K+\Hu^2-\Hu'=\frac{a^2}{2}(\rho+p),\\
    \label{eq:Hus_evol_2}
    &\Hu^2=\frac{a^2}{3}\rho-K.
\end{align}
To split the two epochs we can rewrite the equations as the following:
\begin{align}
\label{eq:KD_evol_a}
    &\Hu'+\frac12(\Hu^2+K)(3c_-^2+1)=\frac{a^2}{2}(c_-^2\rho-p),\\
\label{eq:SR_evol_a}
    &\Hu'-(\Hu^2+K)=-\frac{a^2}{2}(p+\rho).
\end{align}
In this form, we can see that in order to find the scale factor evolution during KD and SR we can set the right-hand side of \cref{eq:KD_evol_a,eq:SR_evol_a}, respectively, to 0. In doing so we arrive at the following:
\begin{align}
  a(\eta) &=
    \begin{cases}
      [S_K(\nu\eta)]^{\frac1\nu} & \eta\in[0,\eta_t]\\
      \frac{[S_K(\nu\eta_t)]^{1+\frac1\nu}}{S_K\left((\nu+1)\eta_t-\eta\right)} & \eta\in[\eta_t,(\nu+1)\eta_t],
    \end{cases}       \\
    \nu&=\frac{3c_-^2+1}{2}
\end{align}
where we define
\begin{equation}
    S_K(x)= \begin{cases}
      \sin{x} & K=+1\\
      x & K=0\\
      \sinh{x} & K=-1.\\
    \end{cases}  
\end{equation}
In the second order differential equations above we implicitly set the two constants of integration by imposing that $a(0)=0$ and setting the overall scaling of the scale factor to 1. The overall scaling has no effect on the analytical expansions, since it only adds a constant to the expansion of $N$. As noted in~\cite{Thavanesan_2020}, when $K=1$, there exists a maximum value for the transition time $\eta_{\text{max}}= \frac{\pi}{2\nu}$, since for values larger than this, the universe starts collapsing prior to reaching the transition. In particular, in this case the value of the curvature density parameter becomes infinite at the onset of inflation. This will be regarded as a breakdown of the approximation.
\subsection{Kinetic dominance}
A pre-inflationary period of kinetic dominance has been introduced in the literature for a standard kinetic Lagrangian $P=X-V(\phi)$. Under broad assumptions it has been shown in \cite{Handley_2014,Hergt_2019} that
classical inflationary universes generically emerge in a
regime where $X \gg V(\phi)$. In this case, any contribution of $\phi$ in the Lagrangian can be dropped. However, using this definition for a general kinetic Lagrangian directly becomes quite restrictive toward the types of models the analytical approximation would allow. For that reason, we will extend the definition of such a regime in the following way. The period of kinetic dominance will be defined as a period of time during which the matter Lagrangian can be approximated to be purely kinetic $P(X,\phi)\to P(X)$. We can see from \cref{eq:p_evol,eq:rho_evol} that in this case the adiabatic sound speed matches the inflationary sound speed $\dot{p}/\dot{\rho}=c_s^2$. In particular, for constant $c_-^2$, this implies the loose description $p/\rho=c_-^2$ used previously.
It has been discussed in~\cite{Unnikrishnan2010} that under the sound speed equality, the scalar field Lagrangian can be redefined to be purely kinetic. Therefore, we can extend the possible permitted Lagrangians to those that can be approximated to be purely kinetic under redefinition: $P(X,\phi)\approx P'(X')$.

Expanding relevant quantities in proper time through logolinear series~\cite{Handley_2019_logo} we have
\begin{align}
    &N=N_p+\frac1\nu\log{\eta}-\frac{K\nu}{6}\eta^2-\frac{K^2\nu^3}{180}\eta^4+\Or(\eta^6),\\
    &\Hu=N'=\frac{1}{\nu\eta}-\frac{K\nu}{3}\eta-\frac{K^2\nu^3}{45}\eta^3+\Or(\eta^5).
\end{align}
where $N_p=\frac1\nu\log{\nu}$. One could also expand $\rho+p$; however, the expansion contains nontrivial noninteger exponents of $\eta$ due to $\nu$ not necessarily being an integer. Despite that, directly expanding the equation of motion yields a rather nice expansion:
\begin{align}
\label{eq:KDexp}
&\left[ \frac{\Z''}{\Z} + \frac{2K}{\Hu}\frac{\Z'}{\Z} + K(3c_-^2-1)\right]=
-\frac{2(3c_-^2-1)}{(3c_-^2+1)^2\eta^2}\nonumber\\ &
\nonumber
+K\left[\frac92c_-^4+16c_-^2+\frac{97}{6}+\frac{\K^2(9c_-^6+36c_-^4+59c_-^2)+96K}{6K(1-c_-^2)-4\K^2c_-^2}\right] \\ &+ \Or(\eta^2).
\end{align}
So for
\begin{align}
    \label{eq:k-def}
    k_-^2=&c_-^2\K^2-K\Bigg[\frac92c_-^4+16c_-^2+\frac{97}{6}+\nonumber\\ &+\frac{\K^2(9c_-^6+36c_-^4+59c_-^2)+96K}{6K(1-c_-^2)-4\K^2c_-^2}\Bigg],
\end{align}
the MS equation of motion becomes 
\begin{equation}
    v_k''+\left[k_-^2+\frac{\nu-1}{\nu^2\eta^2}\right]v_k=0.
\end{equation}
The solution to this can be represented using Hankel functions:
\begin{equation}
v_k=\sqrt{\frac{\pi}{4}}\sqrt{\eta}\left[A_kH_{\frac{1}{2}-\frac{1}{\nu}}^{(1)}(k_-\eta)+B_kH_{\frac{1}{2}-\frac{1}{\nu}}^{(2)}(k_-\eta)\right].    
\end{equation}
Out of quantum quantisation considerations we must pick coefficients such that  $\|B_k\|^2-\|A_k\|^2=1$. Picking the right-handed mode, we set $A_k=0$, $B_k=1$. Note that the question of initial conditions at the start of inflation is still an open problem, with a number of possibilities each having different physical motivations \cite{2021PhRvD.104f3532G}. In this paper the simplest choice is made, which also corresponds to the Bunch-Davies vacuum in the specific case of unit sound speed in a flat universe.
\subsection{Slow roll}
The period of slow roll is formally defined by defining a number of slow-roll parameters and requiring that they remain small during the period, more information on which can be found in  ~\cite{mukhanov1999,Garriga_1999}. In this paper the simplest treatment of slow roll is used by requiring that $\E\to0$.

\begin{figure}
\centering
\includegraphics[width=0.5\textwidth]{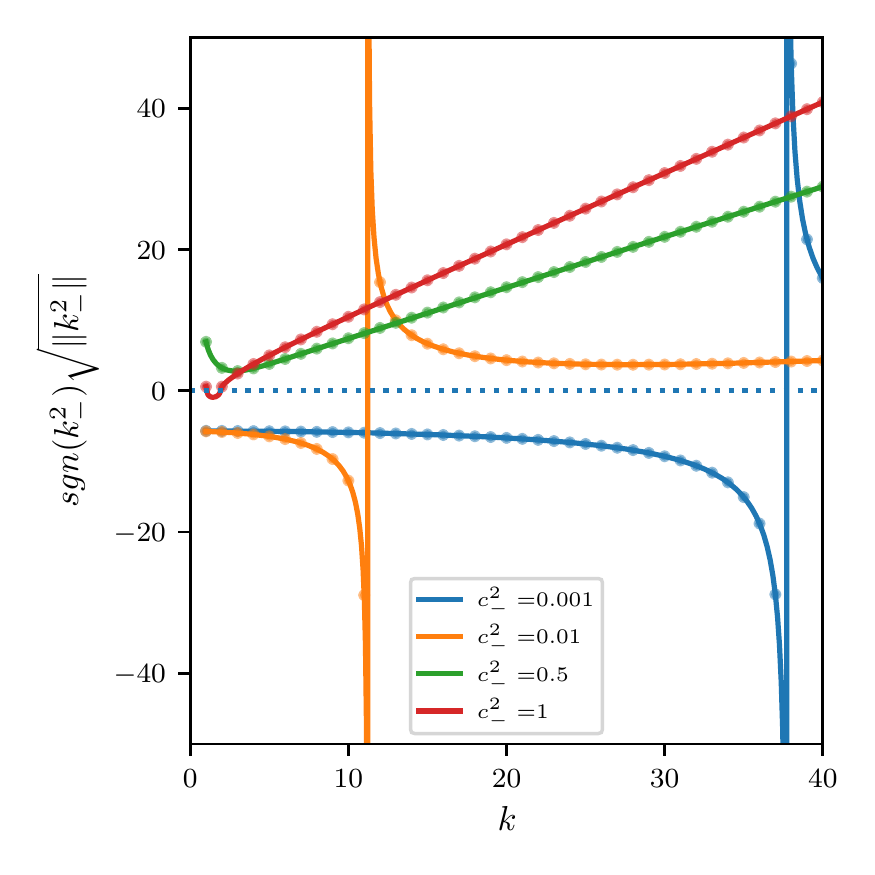}
\caption{\label{fig:k_minus}Plot of $k_-$ for different values of $c_-^2$. As the value of $c_-^2$ decreases, the number of values of integer $k$, which correspond to negative values of $k_-^2$ increases.}
\end{figure}

Now, during SR we can similarly expand the equations of motion to find
\begin{align}
\label{eq:SRexp}
&\left[ \frac{\Z''}{\Z} + \frac{2K}{\Hu}\frac{\Z'}{\Z} + K(3c_+^2-1)\right]\to \frac{a''}{a}+3Kc_+^2
=\\ &\frac{2}{\left((\nu+1)\eta_t-\eta\right)^2}-K\left(\frac13-3c_+^2\right)+\Or(\eta^2)\nonumber
\end{align}
The MS equation of motion becomes
\begin{equation}
    v_k''+\left[k_+^2-\frac{2}{\left(\left(\nu+1\right)\eta_t-\eta\right)^2}\right]v_k=0,
\end{equation}
where 
\begin{equation}
\label{eq:k_plus}
    k_+^2=c_+^2\K^2+K\left(\frac13-3c_+^2\right).
\end{equation}
The solution to this can similarly be represented using Hankel functions:
\begin{align}
v_k=\sqrt{\frac{\pi}{4}}\sqrt{(\nu+1)\eta_t-\eta}\nonumber\bigg[&C_kH_{\frac32}^{(1)}(k_+((\nu+1)\eta_t-\eta))\\+&D_kH_{\frac32}^{(2)}(k_+((\nu+1)\eta_t-\eta))\bigg]    
\end{align}
Comparing to the case of $c_s^2=1$, we see that  values of $k_-^2$ and $k_+^2$ remain finite and positive in the presence of curvature. This is no longer the case when speed of sound is taken into account, which results in different dynamics for different wave numbers.
When the speed of inflation is allowed to be small, from the form of \cref{eq:k-def} we see that near
\begin{align}
    \label{eq:cutoff}
    \K_{\text{cut}}^2=\frac{3K(1-c_-^2)}{2c_-^2},
\end{align}
the effective wave numbers can become negative and extremely large in the case of closed universes ($K = +1$). This can be seen from \cref{fig:k_minus}. In particular all values for wave numbers smaller than this are imaginary. For the rest of the paper we are going to treat this as a natural cutoff for the power spectra for closed universes. It is also important to note, that for specific values of $c_s^2$, the equality in \cref{eq:cutoff} can be satisfied exactly, which corresponds to an infinite constant value in the logolinear expansion. This happens only for a countable number of values of $c_-^2$, since the Fourier decomposition only admits integer wave numbers. Whenever this happens, it should be regarded as a breakdown of the approximation. 

Matching the value of $v_k$ and its first derivative at the transition time, we have the following values for the coefficients:
\begin{align}
    C_k=\frac{i\pi\sqrt{\nu}\eta_t}{4}&\bigg[k_+H_{\frac12}^{(2)}(k_+\nu\eta_t)H_{\frac12-\frac1\nu}^{(2)}(k_-\eta_t)\\\nonumber+&k_-H_{\frac32}^{(2)}(k_+\nu\eta_t)H_{-\frac12-\frac1\nu}^{(2)}(k_-\eta_t)\bigg],
\end{align}
and
\begin{align}
    D_k=-\frac{i\pi\sqrt{\nu}\eta_t}{4}&\bigg[k_+H_{\frac12}^{(1)}(k_+\nu\eta_t)H_{\frac12-\frac1\nu}^{(2)}(k_-\eta_t)\\\nonumber+&k_-H_{\frac32}^{(1)}(k_+\nu\eta_t)H_{-\frac12-\frac1\nu}^{(2)}(k_-\eta_t)\bigg].
\end{align}

\section{Discussion and contact with observations}\label{sec:discussion}
With the approximations in place, we can now calculate the power spectrum of the comoving curvature perturbations at the end of the SR period as 
\begin{align}
\label{eq:PPS}
\PP_{\R}&=\frac{k^3}{2\pi^2}\|\R_k\|^2\nonumber\\
&\to A_s\frac{c_+^3k^3}{k_+^3}\|C_k-D_k\|^2\frac{2}{\frac{c_+}{c_-}+\frac{c_-}{c_+}},
\end{align}
where formally divergent quantities are absorbed into $A_s$. Note, that the overall form for the constant term is chosen using the asymptotic limit of the power spectra for short wavelengths. To be precise, we include an extra harmonic mean term of the two sound speeds due to the limiting behaviour of the $\|C_k-D_k\|^2$ term. 
We can see that in general as $\K^2\to\infty$: $k_+^2\to c_+^2\K^2$ and $k_-^2\to c_-^2\K^2$. So overall from the asymptotic behaviour of Hankel functions
\begin{equation}
    \|C_k-D_k\|^2\to \frac{c_+}{c_-}\sin^2{(k_+\nu\eta_t)}+\frac{c_-}{c_+}\cos^2{(k_+\nu\eta_t)}.
    \label{eq:ckdk}
\end{equation}
In particular, for equal sound speeds in both epochs we recover the standard result of $\PP_{\R}\to A_s$. This is not possible in the case of unequal sound speeds, and the best we can do is to match the mean value $\langle\PP_{\R}\rangle_{k}\to A_s$, which gives us the form in \cref{eq:PPS}.

For the standard K$\Lambda$CDM model, the primordial power spectrum is parametrised by $\PP_{\R}=A_s\left(\frac{k}{k_*}\right)^{n_s-1}$, where $(n_s,A_s)$ are treated as parameters estimated from data. In analogy with ~\cite{Thavanesan_2020}, we will parametrise the full power spectra via 
\begin{equation}
    \PP_{\R}=A_s\left(\frac{k}{k_*}\right)^{n_s-1}\frac{c_+^3k^3}{k_+^3}\|C_k-D_k\|^2\frac{2}{\frac{c_+}{c_-}+\frac{c_-}{c_+}},
\end{equation}
\begin{figure*}
\includegraphics[]{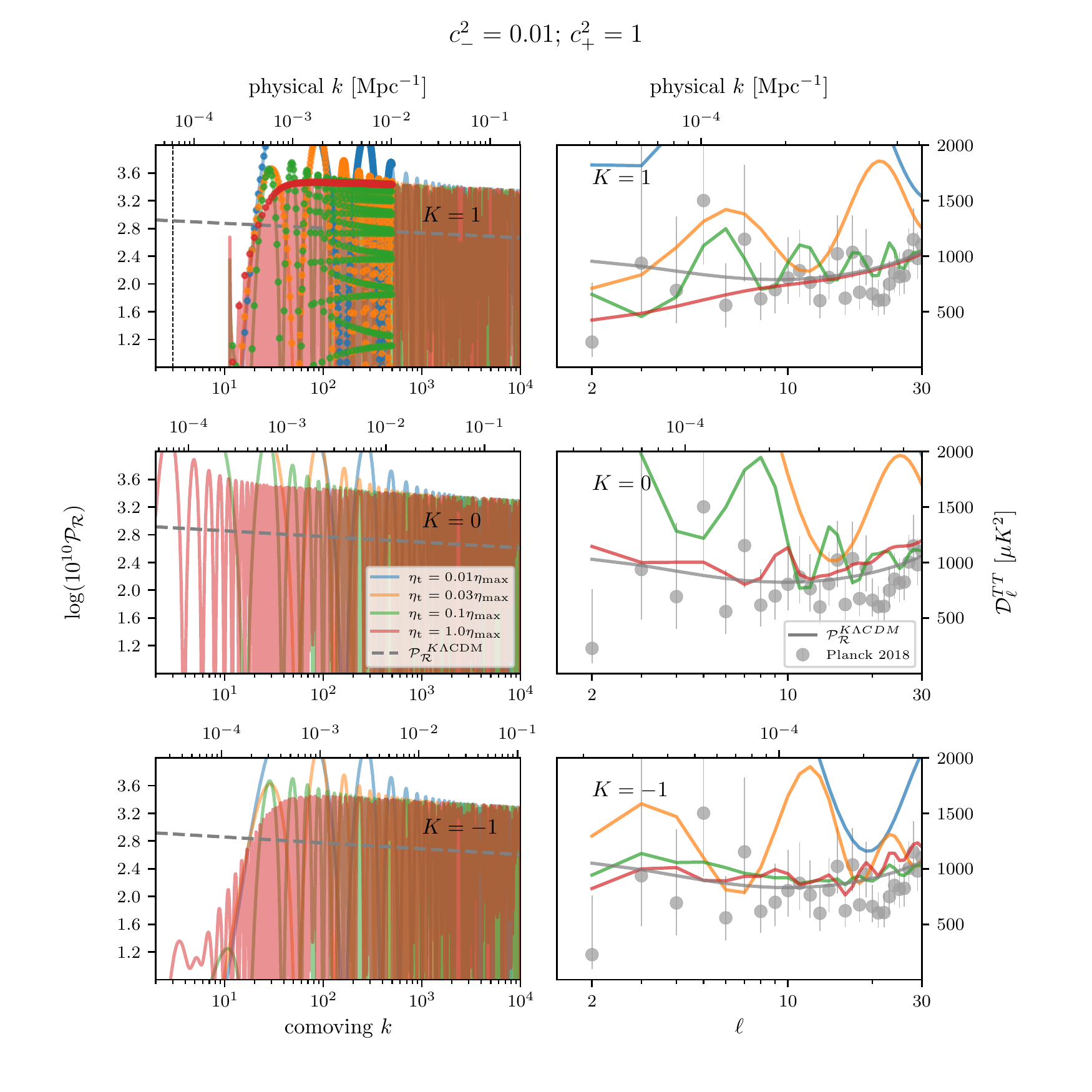}
\caption{\label{fig:small_large}Plot of power spectra generated via the analytical approximation for a variety of different transition times and fixed sound speeds $c_-^2=0.01;c_+^2=1$. For the closed universe the discrete values of wave numbers up to 500 are shown as solid dots and the continuous curves are shown for reference and to ensure consistency with \cite{Thavanesan_2020, Agocs2020}. Shown in grey is the corresponding best fit for the K$\Lambda$CDM model, given a value of $K$.\vspace{50pt}}
\end{figure*}
\begin{figure*}
\includegraphics[]{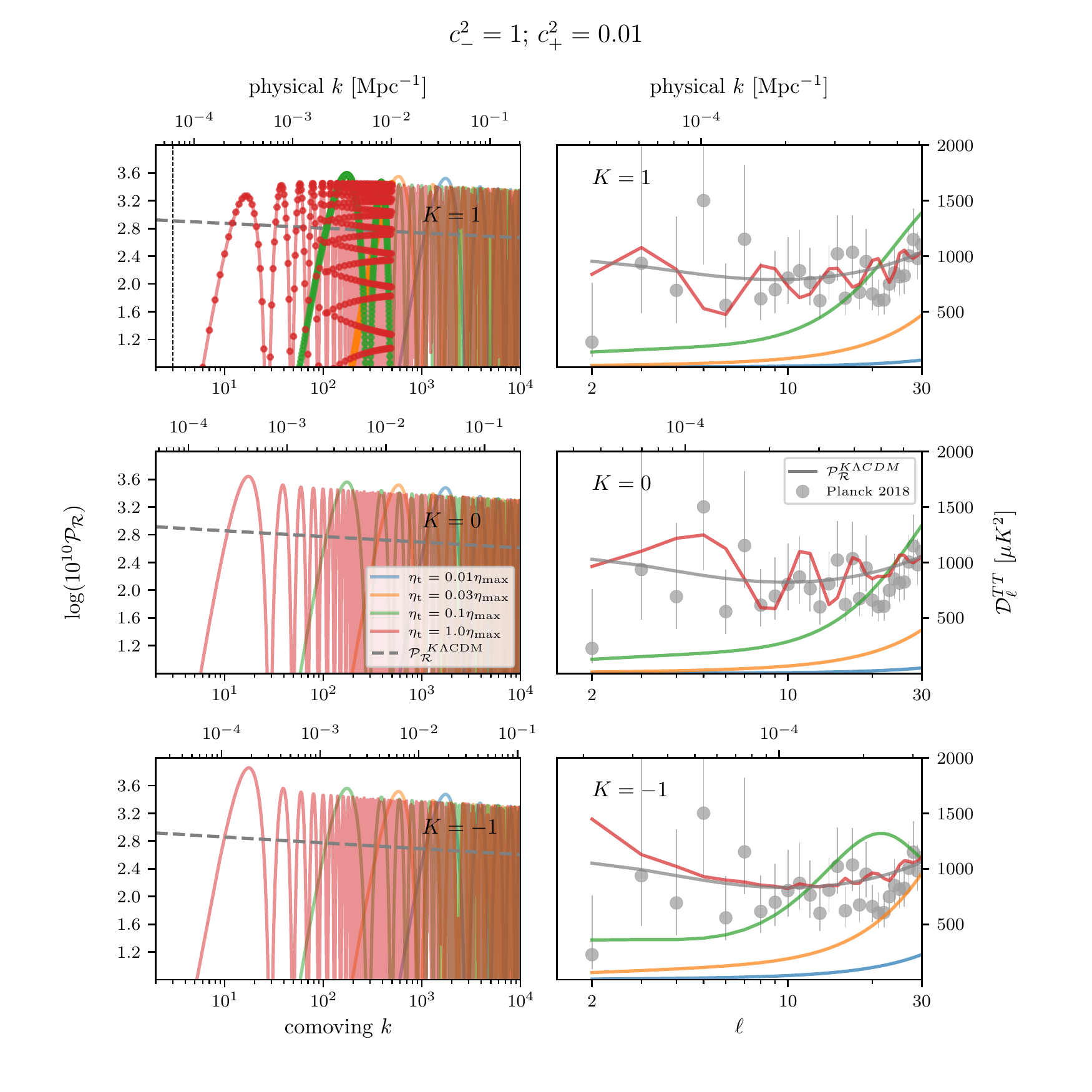}
\caption{\label{fig:large_small}
Plot of power spectra generated via the analytical approximation for a variety of different transition times and fixed sound speeds $c_-^2=1;c_+^2=0.01$. For the closed universe the discrete values of wave numbers up to 500 are shown as solid dots and the continuous curves are shown for reference and to ensure consistency with \cite{Thavanesan_2020, Agocs2020}. Shown in grey is the corresponding best fit for the K$\Lambda$CDM model, given a value of $K$.\vspace{50pt}}
\end{figure*}
where we reintroduce the tilt by hand, since the slow roll approximation is unable to recover it directly.

In summary, curvature produces an overall shift in the dynamical wave vector. However, the introduction of an inflationary sound speed results in nonlinear behaviour of the given shift. The overall effects can be split into two classes, based on the asymptotic behaviour of the resultant power spectra in the limit of large $k$ as shown in \cref{eq:ckdk}.
For that reason, we will separately discuss the effects on the power spectra when the two sound speeds match between the epochs and when they do not.

The CMB spectra created from the primordial power spectra \cite{2011class} are generated using best-fit parameters for each curved case. 
For the closed case we use the Planck 2018 TTTEEE+lowl+lowE+lensing parameters. 
For the flat case we use the $\Lambda$CDM parameters. 
For the open case we calculate the mean posterior distribution of all lensing data using the \texttt{anesthetic} package, subject to
the constraint that $\Omega_K > 0$ \cite{Handley2019anesthetic}. The code for generating figures discussed below is available on request.
\subsection{Nonmatching sound speeds: $c_-\neq c_+$}
We begin by considering the case of nonequal sound speeds. This results in nonvanishing oscillations in both the power spectrum and the angular power spectrum. This violent oscillation is similar to the effect that \citet{2021PhRvD.104f3532G} have observed in the case of ``frozen initial conditions.'' Examples of different sound speeds differing by a factor of 10 in the cases of all possible curved universes are shown in \cref{fig:large_small,fig:small_large}. We see that non-vanishing oscillations persist for both low-$\ell$ and high-$\ell$, and thus can become observable. There are a number of things to note.

First, given a significant increase of sound speed from one epoch to another, the effect of curvature has the opposite effect from the expected one: smaller amounts of curvature significantly promote low-$\ell$ power spectra.
This is a general trend through all possible universe curvatures.

Second, given a significant decrease of sound speed from one epoch to another, the induced oscillations have a smaller frequency, due to the dependence on $k_+$ inside the oscillatory part, while the overall amplitude remains the same.

Third, while differing sound speeds are able to construct and explain a larger number of phenomena, one needs to be careful, as significant jumps will induce oscillations that will be observable in larger multipoles. This may be used to constraint the ratio of the two sound speeds to achieve a better fit to data.

Fourth, due to discretisation of wave numbers in the closed universe, we observe formation of aliased ‘structures’ in the power spectrum, which affect the resulting CMB fits.

\subsection{Matching sound speeds: $c_-=c_+$}
\begin{figure*}
\includegraphics[]{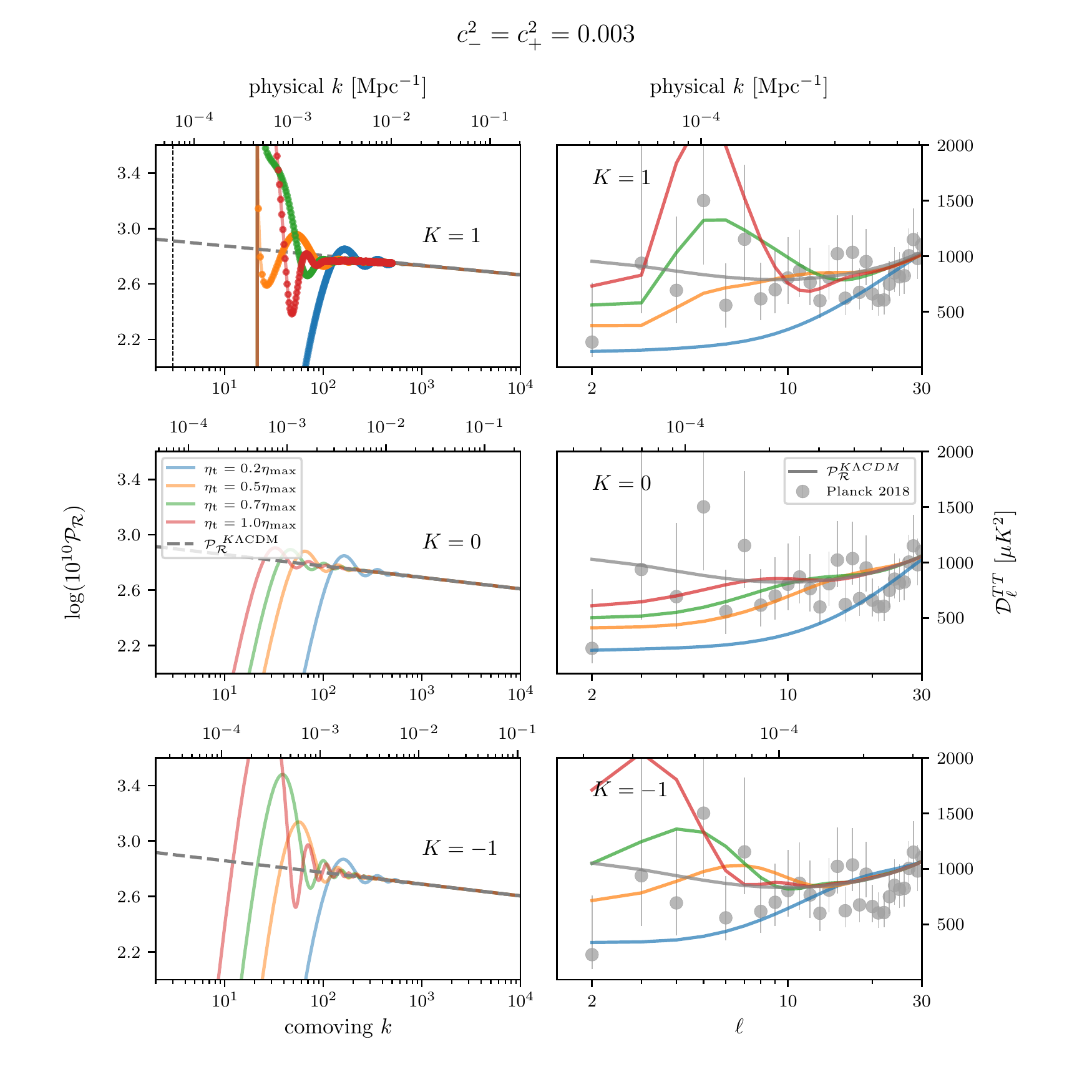}
\caption{Plot of power spectra generated via the analytical approximation for a variety of different transition times and fixed equal sound speed of $c_-^2=c_+^2=0.003$. For the closed universe the discrete values of wave numbers up to 500 are shown as solid dots and the continuous curves are shown for reference and to ensure consistency with \cite{Thavanesan_2020, Agocs2020}. Shown in grey is the corresponding best fit for the K$\Lambda$CDM model, given a value of $K$.\vspace{50pt}}
\label{fig:low_cs}
\end{figure*}

When the two sound speeds match, the calculations and the corresponding behaviours become a lot clearer. An example is shown in \cref{fig:low_cs}. In particular, we observe suppression of power spectra for low wave numbers which is followed by a large power promotion, before converging to the standard K$\Lambda$CDM fit in the limit of a short wavelength. This is observed in both cases of closed and open universes; however, there is no such excitement in the case of flat universes.

\subsection{Fixed transition time}
\begin{figure*}
\includegraphics[]{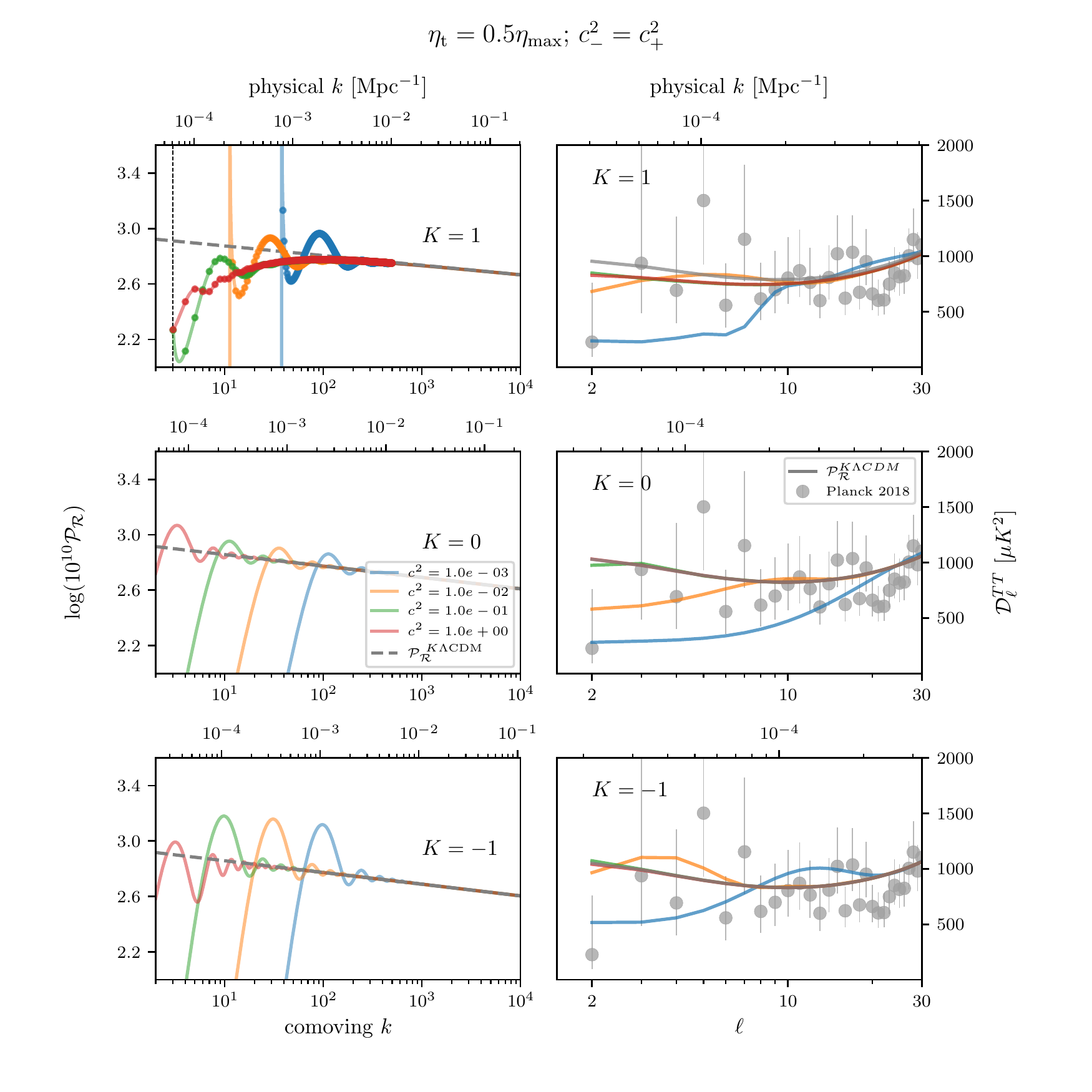}
\caption{Plot of power spectra generated via the analytical approximation for a variety of different sound speeds and fixed transition times. For the closed universe the discrete values of wave numbers up to 500 are shown as solid dots and the continuous curves are shown for reference and to ensure consistency with \cite{Thavanesan_2020, Agocs2020}. Shown in grey is the corresponding best fit for the K$\Lambda$CDM model, given a value of $K$.\vspace{50pt}}
\label{fig:omegak}
\end{figure*}

From a physical point of view it is also of interest for us to compare the effect of $c_s^2$ directly on the amount of primordial curvature, which in this case is parametrised by the transition time $\eta_t$. The overall effect can be seen from \cref{fig:omegak}. We can see that in the case of both flat and closed universes, the angular power spectrum gets significantly suppressed for low $\ell$'s. On the other hand, in the case of open universes, we observe a slight power promotion followed by a similar power suppression. 

\section{Conclusion}\label{sec:conclusion}
In this paper we extended the analytical approximations proposed in \cite{Contaldi_2003, Thavanesan_2020} to analyse the primordial power spectra resulting from a general kinetic Lagrangian. The curved Mukhanov-Sasaki equation and the corresponding action were derived and solved using the analytical approximation with Bunch-Davies initial conditions. 

Through this modification, we arrived at two extra degrees of freedom $c_-$ and $c_+$ on top of the transition time $\eta_t$. These extra parameters could be used to obtain a better fit with data, especially in light of the recent discrepancies that had arisen with the standard $\Lambda$CDM model.

Employing the approximation, we showed that the standard kinetic Lagrangian produced observationally significant alterations to the power spectrum. In the specific case of closed universes we showed that a cutoff in the power spectra was observed due to imaginary wave numbers. Furthermore, a significant change in the inflationary sound speed resulted in nondecaying oscillations. It remains a question of future work to see whether the ringing oscillations persist or a continuous change of sound speed acts to dampen the oscillations when an exact numerical solution is calculated. 

\section*{Acknowledgements}\label{sec:acknowledgements}
WH was supported by a Royal Society University Research Fellowship. ZS was funded through a summer project by the Department of Applied Mathematics and Theoretical Physics (DAMTP) Cambridge Mathematics Placements Programme (CMP).

\bibliography{bibliography.bib}

\begin{thebibliography}{43}%
\makeatletter
\providecommand \@ifxundefined [1]{%
 \@ifx{#1\undefined}
}%
\providecommand \@ifnum [1]{%
 \ifnum #1\expandafter \@firstoftwo
 \else \expandafter \@secondoftwo
 \fi
}%
\providecommand \@ifx [1]{%
 \ifx #1\expandafter \@firstoftwo
 \else \expandafter \@secondoftwo
 \fi
}%
\providecommand \natexlab [1]{#1}%
\providecommand \enquote  [1]{``#1''}%
\providecommand \bibnamefont  [1]{#1}%
\providecommand \bibfnamefont [1]{#1}%
\providecommand \citenamefont [1]{#1}%
\providecommand \href@noop [0]{\@secondoftwo}%
\providecommand \href [0]{\begingroup \@sanitize@url \@href}%
\providecommand \@href[1]{\@@startlink{#1}\@@href}%
\providecommand \@@href[1]{\endgroup#1\@@endlink}%
\providecommand \@sanitize@url [0]{\catcode `\\12\catcode `\$12\catcode
  `\&12\catcode `\#12\catcode `\^12\catcode `\_12\catcode `\%12\relax}%
\providecommand \@@startlink[1]{}%
\providecommand \@@endlink[0]{}%
\providecommand \url  [0]{\begingroup\@sanitize@url \@url }%
\providecommand \@url [1]{\endgroup\@href {#1}{\urlprefix }}%
\providecommand \urlprefix  [0]{URL }%
\providecommand \Eprint [0]{\href }%
\providecommand \doibase [0]{https://doi.org/}%
\providecommand \selectlanguage [0]{\@gobble}%
\providecommand \bibinfo  [0]{\@secondoftwo}%
\providecommand \bibfield  [0]{\@secondoftwo}%
\providecommand \translation [1]{[#1]}%
\providecommand \BibitemOpen [0]{}%
\providecommand \bibitemStop [0]{}%
\providecommand \bibitemNoStop [0]{.\EOS\space}%
\providecommand \EOS [0]{\spacefactor3000\relax}%
\providecommand \BibitemShut  [1]{\csname bibitem#1\endcsname}%
\let\auto@bib@innerbib\@empty
\bibitem [{\citenamefont {Contaldi}\ \emph {et~al.}(2003)\citenamefont
  {Contaldi}, \citenamefont {Peloso}, \citenamefont {Kofman},\ and\
  \citenamefont {Linde}}]{Contaldi_2003}%
  \BibitemOpen
  \bibfield  {author} {\bibinfo {author} {\bibfnamefont {C.~R.}\ \bibnamefont
  {Contaldi}}, \bibinfo {author} {\bibfnamefont {M.}~\bibnamefont {Peloso}},
  \bibinfo {author} {\bibfnamefont {L.}~\bibnamefont {Kofman}},\ and\ \bibinfo
  {author} {\bibfnamefont {A.}~\bibnamefont {Linde}},\ }\bibfield  {title}
  {\bibinfo {title} {Suppressing the lower multipoles in the cmb
  anisotropies},\ }\href {https://doi.org/10.1088/1475-7516/2003/07/002}
  {\bibfield  {journal} {\bibinfo  {journal} {Journal of Cosmology and
  Astroparticle Physics}\ }\textbf {\bibinfo {volume} {2003}}\bibinfo  {number}
  { (07)},\ \bibinfo {pages} {002–002}}\BibitemShut {NoStop}%
\bibitem [{\citenamefont {Thavanesan}\ \emph {et~al.}(2021)\citenamefont
  {Thavanesan}, \citenamefont {Werth},\ and\ \citenamefont
  {Handley}}]{Thavanesan_2020}%
  \BibitemOpen
\bibfield  {number} {  }\bibfield  {author} {\bibinfo {author} {\bibfnamefont
  {A.}~\bibnamefont {Thavanesan}}, \bibinfo {author} {\bibfnamefont
  {D.}~\bibnamefont {Werth}},\ and\ \bibinfo {author} {\bibfnamefont
  {W.}~\bibnamefont {Handley}},\ }\bibfield  {title} {\bibinfo {title}
  {Analytical approximations for curved primordial power spectra},\ }\href
  {https://doi.org/10.1103/PhysRevD.103.023519} {\bibfield  {journal} {\bibinfo
   {journal} {Phys. Rev. D}\ }\textbf {\bibinfo {volume} {103}},\ \bibinfo
  {pages} {023519} (\bibinfo {year} {2021})}\BibitemShut {NoStop}%
\bibitem [{\citenamefont {Starobinsky}(1979)}]{Starobinskii1979}%
  \BibitemOpen
  \bibfield  {author} {\bibinfo {author} {\bibfnamefont {A.~A.}\ \bibnamefont
  {Starobinsky}},\ }\bibfield  {title} {\bibinfo {title} {{Spectrum of relict
  gravitational radiation and the early state of the universe}},\ }\href@noop
  {} {\bibfield  {journal} {\bibinfo  {journal} {JETP Lett.}\ }\textbf
  {\bibinfo {volume} {30}},\ \bibinfo {pages} {682} (\bibinfo {year}
  {1979})}\BibitemShut {NoStop}%
\bibitem [{\citenamefont {{Guth}}(1981)}]{Guth1981}%
  \BibitemOpen
  \bibfield  {author} {\bibinfo {author} {\bibfnamefont {A.~H.}\ \bibnamefont
  {{Guth}}},\ }\bibfield  {title} {\bibinfo {title} {{Inflationary universe: A
  possible solution to the horizon and flatness problems}},\ }\href
  {https://doi.org/10.1103/PhysRevD.23.347} {\bibfield  {journal} {\bibinfo
  {journal} {\prd}\ }\textbf {\bibinfo {volume} {23}},\ \bibinfo {pages} {347}
  (\bibinfo {year} {1981})}\BibitemShut {NoStop}%
\bibitem [{\citenamefont {{Linde}}(1982)}]{Linde1982}%
  \BibitemOpen
  \bibfield  {author} {\bibinfo {author} {\bibfnamefont {A.~D.}\ \bibnamefont
  {{Linde}}},\ }\bibfield  {title} {\bibinfo {title} {{A new inflationary
  universe scenario: A possible solution of the horizon, flatness, homogeneity,
  isotropy and primordial monopole problems}},\ }\href
  {https://doi.org/10.1016/0370-2693(82)91219-9} {\bibfield  {journal}
  {\bibinfo  {journal} {Physics Letters B}\ }\textbf {\bibinfo {volume}
  {108}},\ \bibinfo {pages} {389} (\bibinfo {year} {1982})}\BibitemShut
  {NoStop}%
\bibitem [{\citenamefont {{Planck
  Collaboration}}(2018{\natexlab{a}})}]{planck_parameters}%
  \BibitemOpen
  \bibfield  {author} {\bibinfo {author} {\bibnamefont {{Planck
  Collaboration}}},\ }\bibfield  {title} {\bibinfo {title} {{Planck 2018
  results. VI. Cosmological parameters}},\ }\href@noop {} {\bibfield  {journal}
  {\bibinfo  {journal} {arXiv e-prints}\ ,\ \bibinfo {eid} {arXiv:1807.06209}}
  (\bibinfo {year} {2018}{\natexlab{a}})},\ \Eprint
  {https://arxiv.org/abs/1807.06209} {arXiv:1807.06209 [astro-ph.CO]}
  \BibitemShut {NoStop}%
\bibitem [{\citenamefont {{Planck
  Collaboration}}(2018{\natexlab{b}})}]{planck_inflation2018}%
  \BibitemOpen
  \bibfield  {author} {\bibinfo {author} {\bibnamefont {{Planck
  Collaboration}}},\ }\bibfield  {title} {\bibinfo {title} {{Planck 2018
  results. X. Constraints on inflation}},\ }\href@noop {} {\bibfield  {journal}
  {\bibinfo  {journal} {arXiv e-prints}\ ,\ \bibinfo {eid} {arXiv:1807.06211}}
  (\bibinfo {year} {2018}{\natexlab{b}})},\ \Eprint
  {https://arxiv.org/abs/1807.06211} {arXiv:1807.06211 [astro-ph.CO]}
  \BibitemShut {NoStop}%
\bibitem [{\citenamefont {{Handley}}\ \emph {et~al.}(2019)\citenamefont
  {{Handley}}, \citenamefont {{Lasenby}}, \citenamefont {{Peiris}},\ and\
  \citenamefont {{Hobson}}}]{2019PhRvD.100j3511H}%
  \BibitemOpen
  \bibfield  {author} {\bibinfo {author} {\bibfnamefont {W.~J.}\ \bibnamefont
  {{Handley}}}, \bibinfo {author} {\bibfnamefont {A.~N.}\ \bibnamefont
  {{Lasenby}}}, \bibinfo {author} {\bibfnamefont {H.~V.}\ \bibnamefont
  {{Peiris}}},\ and\ \bibinfo {author} {\bibfnamefont {M.~P.}\ \bibnamefont
  {{Hobson}}},\ }\bibfield  {title} {\bibinfo {title} {{Bayesian inflationary
  reconstructions from Planck 2018 data}},\ }\href
  {https://doi.org/10.1103/PhysRevD.100.103511} {\bibfield  {journal} {\bibinfo
   {journal} {\prd}\ }\textbf {\bibinfo {volume} {100}},\ \bibinfo {eid}
  {103511} (\bibinfo {year} {2019})},\ \Eprint
  {https://arxiv.org/abs/1908.00906} {arXiv:1908.00906 [astro-ph.CO]}
  \BibitemShut {NoStop}%
\bibitem [{\citenamefont {{Handley}}\ \emph {et~al.}(2014)\citenamefont
  {{Handley}}, \citenamefont {{Brechet}}, \citenamefont {{Lasenby}},\ and\
  \citenamefont {{Hobson}}}]{kinetic_dominance}%
  \BibitemOpen
  \bibfield  {author} {\bibinfo {author} {\bibfnamefont {W.~J.}\ \bibnamefont
  {{Handley}}}, \bibinfo {author} {\bibfnamefont {S.~D.}\ \bibnamefont
  {{Brechet}}}, \bibinfo {author} {\bibfnamefont {A.~N.}\ \bibnamefont
  {{Lasenby}}},\ and\ \bibinfo {author} {\bibfnamefont {M.~P.}\ \bibnamefont
  {{Hobson}}},\ }\bibfield  {title} {\bibinfo {title} {{Kinetic initial
  conditions for inflation}},\ }\href
  {https://doi.org/10.1103/PhysRevD.89.063505} {\bibfield  {journal} {\bibinfo
  {journal} {\prd}\ }\textbf {\bibinfo {volume} {89}},\ \bibinfo {eid} {063505}
  (\bibinfo {year} {2014})},\ \Eprint {https://arxiv.org/abs/1401.2253}
  {arXiv:1401.2253} \BibitemShut {NoStop}%
\bibitem [{\citenamefont {{Hergt}}\ \emph
  {et~al.}(2018{\natexlab{a}})\citenamefont {{Hergt}}, \citenamefont
  {{Handley}}, \citenamefont {{Hobson}},\ and\ \citenamefont
  {{Lasenby}}}]{Hergt1}%
  \BibitemOpen
  \bibfield  {author} {\bibinfo {author} {\bibfnamefont {L.~T.}\ \bibnamefont
  {{Hergt}}}, \bibinfo {author} {\bibfnamefont {W.~J.}\ \bibnamefont
  {{Handley}}}, \bibinfo {author} {\bibfnamefont {M.~P.}\ \bibnamefont
  {{Hobson}}},\ and\ \bibinfo {author} {\bibfnamefont {A.~N.}\ \bibnamefont
  {{Lasenby}}},\ }\bibfield  {title} {\bibinfo {title} {{A case for kinetically
  dominated initial conditions for inflation}},\ }\href@noop {} {\bibfield
  {journal} {\bibinfo  {journal} {ArXiv e-prints}\ } (\bibinfo {year}
  {2018}{\natexlab{a}})},\ \Eprint {https://arxiv.org/abs/1809.07185}
  {arXiv:1809.07185} \BibitemShut {NoStop}%
\bibitem [{\citenamefont {Schwarz}\ and\ \citenamefont
  {Ramirez}(2009)}]{just_enough_inflation}%
  \BibitemOpen
  \bibfield  {author} {\bibinfo {author} {\bibfnamefont {D.~J.}\ \bibnamefont
  {Schwarz}}\ and\ \bibinfo {author} {\bibfnamefont {E.}~\bibnamefont
  {Ramirez}},\ }\bibfield  {title} {\bibinfo {title} {{Just enough
  inflation}},\ }in\ \href {https://doi.org/10.1142/9789814374552_0180} {\emph
  {\bibinfo {booktitle} {{12th Marcel Grossmann Meeting on General
  Relativity}}}}\ (\bibinfo {year} {2009})\ pp.\ \bibinfo {pages}
  {1241--1243},\ \Eprint {https://arxiv.org/abs/0912.4348} {arXiv:0912.4348
  [hep-ph]} \BibitemShut {NoStop}%
\bibitem [{\citenamefont {{Cicoli}}\ \emph {et~al.}(2014)\citenamefont
  {{Cicoli}}, \citenamefont {{Downes}}, \citenamefont {{Dutta}}, \citenamefont
  {{Pedro}},\ and\ \citenamefont {{Westphal}}}]{just_enough_inflation2}%
  \BibitemOpen
  \bibfield  {author} {\bibinfo {author} {\bibfnamefont {M.}~\bibnamefont
  {{Cicoli}}}, \bibinfo {author} {\bibfnamefont {S.}~\bibnamefont {{Downes}}},
  \bibinfo {author} {\bibfnamefont {B.}~\bibnamefont {{Dutta}}}, \bibinfo
  {author} {\bibfnamefont {F.~G.}\ \bibnamefont {{Pedro}}},\ and\ \bibinfo
  {author} {\bibfnamefont {A.}~\bibnamefont {{Westphal}}},\ }\bibfield  {title}
  {\bibinfo {title} {{Just enough inflation: power spectrum modifications at
  large scales}},\ }\href {https://doi.org/10.1088/1475-7516/2014/12/030}
  {\bibfield  {journal} {\bibinfo  {journal} {\jcap}\ }\textbf {\bibinfo
  {volume} {2014}},\ \bibinfo {eid} {030} (\bibinfo {year} {2014})},\ \Eprint
  {https://arxiv.org/abs/1407.1048} {arXiv:1407.1048 [hep-th]} \BibitemShut
  {NoStop}%
\bibitem [{\citenamefont {{Boyanovsky}}\ \emph
  {et~al.}(2006{\natexlab{a}})\citenamefont {{Boyanovsky}}, \citenamefont {{de
  Vega}},\ and\ \citenamefont {{Sanchez}}}]{BVS1}%
  \BibitemOpen
  \bibfield  {author} {\bibinfo {author} {\bibfnamefont {D.}~\bibnamefont
  {{Boyanovsky}}}, \bibinfo {author} {\bibfnamefont {H.~J.}\ \bibnamefont {{de
  Vega}}},\ and\ \bibinfo {author} {\bibfnamefont {N.~G.}\ \bibnamefont
  {{Sanchez}}},\ }\bibfield  {title} {\bibinfo {title} {{CMB quadrupole
  suppression. I. Initial conditions of inflationary perturbations}},\ }\href
  {https://doi.org/10.1103/PhysRevD.74.123006} {\bibfield  {journal} {\bibinfo
  {journal} {\prd}\ }\textbf {\bibinfo {volume} {74}},\ \bibinfo {eid} {123006}
  (\bibinfo {year} {2006}{\natexlab{a}})},\ \Eprint
  {https://arxiv.org/abs/astro-ph/0607508} {arXiv:astro-ph/0607508 [astro-ph]}
  \BibitemShut {NoStop}%
\bibitem [{\citenamefont {{Boyanovsky}}\ \emph
  {et~al.}(2006{\natexlab{b}})\citenamefont {{Boyanovsky}}, \citenamefont {{de
  Vega}},\ and\ \citenamefont {{Sanchez}}}]{BVS2}%
  \BibitemOpen
  \bibfield  {author} {\bibinfo {author} {\bibfnamefont {D.}~\bibnamefont
  {{Boyanovsky}}}, \bibinfo {author} {\bibfnamefont {H.~J.}\ \bibnamefont {{de
  Vega}}},\ and\ \bibinfo {author} {\bibfnamefont {N.~G.}\ \bibnamefont
  {{Sanchez}}},\ }\bibfield  {title} {\bibinfo {title} {{CMB quadrupole
  suppression. II. The early fast roll stage}},\ }\href
  {https://doi.org/10.1103/PhysRevD.74.123007} {\bibfield  {journal} {\bibinfo
  {journal} {\prd}\ }\textbf {\bibinfo {volume} {74}},\ \bibinfo {eid} {123007}
  (\bibinfo {year} {2006}{\natexlab{b}})},\ \Eprint
  {https://arxiv.org/abs/astro-ph/0607487} {arXiv:astro-ph/0607487 [astro-ph]}
  \BibitemShut {NoStop}%
\bibitem [{\citenamefont {Avis}\ \emph {et~al.}(2020)\citenamefont {Avis},
  \citenamefont {Jazayeri}, \citenamefont {Pajer},\ and\ \citenamefont
  {Supeł}}]{Avis_2020}%
  \BibitemOpen
  \bibfield  {author} {\bibinfo {author} {\bibfnamefont {G.}~\bibnamefont
  {Avis}}, \bibinfo {author} {\bibfnamefont {S.}~\bibnamefont {Jazayeri}},
  \bibinfo {author} {\bibfnamefont {E.}~\bibnamefont {Pajer}},\ and\ \bibinfo
  {author} {\bibfnamefont {J.}~\bibnamefont {Supeł}},\ }\bibfield  {title}
  {\bibinfo {title} {Spatial curvature at the sound horizon},\ }\href
  {https://doi.org/10.1088/1475-7516/2020/02/034} {\bibfield  {journal}
  {\bibinfo  {journal} {Journal of Cosmology and Astroparticle Physics}\
  }\textbf {\bibinfo {volume} {2020}}\bibinfo  {number} { (02)},\ \bibinfo
  {pages} {034–034}}\BibitemShut {NoStop}%
\bibitem [{\citenamefont {{Planck Collaboration}}(2019)}]{planck_isotropy}%
  \BibitemOpen
\bibfield  {number} {  }\bibfield  {author} {\bibinfo {author} {\bibnamefont
  {{Planck Collaboration}}},\ }\bibfield  {title} {\bibinfo {title} {{Planck
  2018 results. VII. Isotropy and Statistics of the CMB}},\ }\href@noop {}
  {\bibfield  {journal} {\bibinfo  {journal} {arXiv e-prints}\ ,\ \bibinfo
  {eid} {arXiv:1906.02552}} (\bibinfo {year} {2019})},\ \Eprint
  {https://arxiv.org/abs/1906.02552} {arXiv:1906.02552 [astro-ph.CO]}
  \BibitemShut {NoStop}%
\bibitem [{\citenamefont {Handley}(2021)}]{2019arXiv190809139H}%
  \BibitemOpen
  \bibfield  {author} {\bibinfo {author} {\bibfnamefont {W.}~\bibnamefont
  {Handley}},\ }\bibfield  {title} {\bibinfo {title} {{Curvature tension:
  evidence for a closed universe}},\ }\href
  {https://doi.org/10.1103/PhysRevD.103.L041301} {\bibfield  {journal}
  {\bibinfo  {journal} {Phys. Rev. D}\ }\textbf {\bibinfo {volume} {103}},\
  \bibinfo {pages} {L041301} (\bibinfo {year} {2021})},\ \Eprint
  {https://arxiv.org/abs/1908.09139} {arXiv:1908.09139 [astro-ph.CO]}
  \BibitemShut {NoStop}%
\bibitem [{\citenamefont {{Di Valentino}}\ \emph {et~al.}(2020)\citenamefont
  {{Di Valentino}}, \citenamefont {{Melchiorri}},\ and\ \citenamefont
  {{Silk}}}]{2020NatAs...4..196D}%
  \BibitemOpen
  \bibfield  {author} {\bibinfo {author} {\bibfnamefont {E.}~\bibnamefont {{Di
  Valentino}}}, \bibinfo {author} {\bibfnamefont {A.}~\bibnamefont
  {{Melchiorri}}},\ and\ \bibinfo {author} {\bibfnamefont {J.}~\bibnamefont
  {{Silk}}},\ }\bibfield  {title} {\bibinfo {title} {{Planck evidence for a
  closed Universe and a possible crisis for cosmology}},\ }\href
  {https://doi.org/10.1038/s41550-019-0906-9} {\bibfield  {journal} {\bibinfo
  {journal} {Nature Astronomy}\ }\textbf {\bibinfo {volume} {4}},\ \bibinfo
  {pages} {196} (\bibinfo {year} {2020})},\ \Eprint
  {https://arxiv.org/abs/1911.02087} {arXiv:1911.02087 [astro-ph.CO]}
  \BibitemShut {NoStop}%
\bibitem [{\citenamefont {{Efstathiou}}\ and\ \citenamefont
  {{Gratton}}(2020)}]{2020MNRAS.496L..91E}%
  \BibitemOpen
  \bibfield  {author} {\bibinfo {author} {\bibfnamefont {G.}~\bibnamefont
  {{Efstathiou}}}\ and\ \bibinfo {author} {\bibfnamefont {S.}~\bibnamefont
  {{Gratton}}},\ }\bibfield  {title} {\bibinfo {title} {{The evidence for a
  spatially flat Universe}},\ }\href {https://doi.org/10.1093/mnrasl/slaa093}
  {\bibfield  {journal} {\bibinfo  {journal} {\mnras}\ }\textbf {\bibinfo
  {volume} {496}},\ \bibinfo {pages} {L91} (\bibinfo {year} {2020})},\ \Eprint
  {https://arxiv.org/abs/2002.06892} {arXiv:2002.06892 [astro-ph.CO]}
  \BibitemShut {NoStop}%
\bibitem [{\citenamefont {Ellis}\ and\ \citenamefont
  {Maartens}(2003)}]{Ellis_2003}%
  \BibitemOpen
  \bibfield  {author} {\bibinfo {author} {\bibfnamefont {G.~F.~R.}\
  \bibnamefont {Ellis}}\ and\ \bibinfo {author} {\bibfnamefont
  {R.}~\bibnamefont {Maartens}},\ }\bibfield  {title} {\bibinfo {title} {The
  emergent universe: inflationary cosmology with no singularity},\ }\href
  {https://doi.org/10.1088/0264-9381/21/1/015} {\bibfield  {journal} {\bibinfo
  {journal} {Classical and Quantum Gravity}\ }\textbf {\bibinfo {volume}
  {21}},\ \bibinfo {pages} {223–232} (\bibinfo {year} {2003})}\BibitemShut
  {NoStop}%
\bibitem [{\citenamefont {{Lasenby}}\ and\ \citenamefont
  {{Doran}}(2005)}]{lasenbyclosed}%
  \BibitemOpen
  \bibfield  {author} {\bibinfo {author} {\bibfnamefont {A.}~\bibnamefont
  {{Lasenby}}}\ and\ \bibinfo {author} {\bibfnamefont {C.}~\bibnamefont
  {{Doran}}},\ }\bibfield  {title} {\bibinfo {title} {{Closed universes, de
  Sitter space, and inflation}},\ }\href
  {https://doi.org/10.1103/PhysRevD.71.063502} {\bibfield  {journal} {\bibinfo
  {journal} {\prd}\ }\textbf {\bibinfo {volume} {71}},\ \bibinfo {eid} {063502}
  (\bibinfo {year} {2005})},\ \Eprint {https://arxiv.org/abs/astro-ph/0307311}
  {astro-ph/0307311} \BibitemShut {NoStop}%
\bibitem [{\citenamefont {{Verde}}\ \emph {et~al.}(2019)\citenamefont
  {{Verde}}, \citenamefont {{Treu}},\ and\ \citenamefont
  {{Riess}}}]{2019NatAs...3..891V}%
  \BibitemOpen
  \bibfield  {author} {\bibinfo {author} {\bibfnamefont {L.}~\bibnamefont
  {{Verde}}}, \bibinfo {author} {\bibfnamefont {T.}~\bibnamefont {{Treu}}},\
  and\ \bibinfo {author} {\bibfnamefont {A.~G.}\ \bibnamefont {{Riess}}},\
  }\bibfield  {title} {\bibinfo {title} {{Tensions between the early and late
  Universe}},\ }\href {https://doi.org/10.1038/s41550-019-0902-0} {\bibfield
  {journal} {\bibinfo  {journal} {Nature Astronomy}\ }\textbf {\bibinfo
  {volume} {3}},\ \bibinfo {pages} {891} (\bibinfo {year} {2019})},\ \Eprint
  {https://arxiv.org/abs/1907.10625} {arXiv:1907.10625 [astro-ph.CO]}
  \BibitemShut {NoStop}%
\bibitem [{\citenamefont {{Di Valentino}}\ \emph {et~al.}(2021)\citenamefont
  {{Di Valentino}}, \citenamefont {{Mena}}, \citenamefont {{Pan}},
  \citenamefont {{Visinelli}}, \citenamefont {{Yang}}, \citenamefont
  {{Melchiorri}}, \citenamefont {{Mota}}, \citenamefont {{Riess}},\ and\
  \citenamefont {{Silk}}}]{2021CQGra..38o3001D}%
  \BibitemOpen
  \bibfield  {author} {\bibinfo {author} {\bibfnamefont {E.}~\bibnamefont {{Di
  Valentino}}}, \bibinfo {author} {\bibfnamefont {O.}~\bibnamefont {{Mena}}},
  \bibinfo {author} {\bibfnamefont {S.}~\bibnamefont {{Pan}}}, \bibinfo
  {author} {\bibfnamefont {L.}~\bibnamefont {{Visinelli}}}, \bibinfo {author}
  {\bibfnamefont {W.}~\bibnamefont {{Yang}}}, \bibinfo {author} {\bibfnamefont
  {A.}~\bibnamefont {{Melchiorri}}}, \bibinfo {author} {\bibfnamefont {D.~F.}\
  \bibnamefont {{Mota}}}, \bibinfo {author} {\bibfnamefont {A.~G.}\
  \bibnamefont {{Riess}}},\ and\ \bibinfo {author} {\bibfnamefont
  {J.}~\bibnamefont {{Silk}}},\ }\bibfield  {title} {\bibinfo {title} {{In the
  realm of the Hubble tension-a review of solutions}},\ }\href
  {https://doi.org/10.1088/1361-6382/ac086d} {\bibfield  {journal} {\bibinfo
  {journal} {Classical and Quantum Gravity}\ }\textbf {\bibinfo {volume}
  {38}},\ \bibinfo {eid} {153001} (\bibinfo {year} {2021})},\ \Eprint
  {https://arxiv.org/abs/2103.01183} {arXiv:2103.01183 [astro-ph.CO]}
  \BibitemShut {NoStop}%
\bibitem [{\citenamefont {{Hergt}}\ \emph
  {et~al.}(2018{\natexlab{b}})\citenamefont {{Hergt}}, \citenamefont
  {{Handley}}, \citenamefont {{Hobson}},\ and\ \citenamefont
  {{Lasenby}}}]{Hergt2}%
  \BibitemOpen
  \bibfield  {author} {\bibinfo {author} {\bibfnamefont {L.~T.}\ \bibnamefont
  {{Hergt}}}, \bibinfo {author} {\bibfnamefont {W.~J.}\ \bibnamefont
  {{Handley}}}, \bibinfo {author} {\bibfnamefont {M.~P.}\ \bibnamefont
  {{Hobson}}},\ and\ \bibinfo {author} {\bibfnamefont {A.~N.}\ \bibnamefont
  {{Lasenby}}},\ }\bibfield  {title} {\bibinfo {title} {{Constraining the
  kinetically dominated Universe}},\ }\href@noop {} {\bibfield  {journal}
  {\bibinfo  {journal} {ArXiv e-prints}\ } (\bibinfo {year}
  {2018}{\natexlab{b}})},\ \Eprint {https://arxiv.org/abs/1809.07737}
  {arXiv:1809.07737} \BibitemShut {NoStop}%
\bibitem [{\citenamefont {{Hergt}}\ \emph {et~al.}(2020)\citenamefont
  {{Hergt}}, \citenamefont {{Agocs}}, \citenamefont {{Handley}}, \citenamefont
  {{Hobson}},\ and\ \citenamefont {{Lasenby}}}]{Lukas_2020}%
  \BibitemOpen
  \bibfield  {author} {\bibinfo {author} {\bibfnamefont {L.~T.}\ \bibnamefont
  {{Hergt}}}, \bibinfo {author} {\bibfnamefont {F.~J.}\ \bibnamefont
  {{Agocs}}}, \bibinfo {author} {\bibfnamefont {W.~J.}\ \bibnamefont
  {{Handley}}}, \bibinfo {author} {\bibfnamefont {M.~P.}\ \bibnamefont
  {{Hobson}}},\ and\ \bibinfo {author} {\bibfnamefont {A.~N.}\ \bibnamefont
  {{Lasenby}}},\ }\bibfield  {title} {\bibinfo {title} {{Finite inflation in
  curved space}},\ }\href@noop {} {\bibfield  {journal} {\bibinfo  {journal}
  {arXiv e-prints}\ ,\ \bibinfo {eid} {arXiv:2012.xxxxx}} (\bibinfo {year}
  {2020})},\ \Eprint {https://arxiv.org/abs/2012.xxxxx} {arXiv:2012.xxxxx
  [astro-ph.CO]} \BibitemShut {NoStop}%
\bibitem [{\citenamefont {Handley}(2019{\natexlab{a}})}]{Handley_2019}%
  \BibitemOpen
  \bibfield  {author} {\bibinfo {author} {\bibfnamefont {W.}~\bibnamefont
  {Handley}},\ }\bibfield  {title} {\bibinfo {title} {{Primordial power spectra
  for curved inflating universes}},\ }\href
  {https://doi.org/10.1103/PhysRevD.100.123517} {\bibfield  {journal} {\bibinfo
   {journal} {Phys. Rev. D}\ }\textbf {\bibinfo {volume} {100}},\ \bibinfo
  {pages} {123517} (\bibinfo {year} {2019}{\natexlab{a}})},\ \Eprint
  {https://arxiv.org/abs/1907.08524} {arXiv:1907.08524 [astro-ph.CO]}
  \BibitemShut {NoStop}%
\bibitem [{\citenamefont {{Gessey-Jones}}\ and\ \citenamefont
  {{Handley}}(2021)}]{2021PhRvD.104f3532G}%
  \BibitemOpen
  \bibfield  {author} {\bibinfo {author} {\bibfnamefont {T.}~\bibnamefont
  {{Gessey-Jones}}}\ and\ \bibinfo {author} {\bibfnamefont {W.~J.}\
  \bibnamefont {{Handley}}},\ }\bibfield  {title} {\bibinfo {title}
  {{Constraining quantum initial conditions before inflation}},\ }\href
  {https://doi.org/10.1103/PhysRevD.104.063532} {\bibfield  {journal} {\bibinfo
   {journal} {\prd}\ }\textbf {\bibinfo {volume} {104}},\ \bibinfo {eid}
  {063532} (\bibinfo {year} {2021})},\ \Eprint
  {https://arxiv.org/abs/2104.03016} {arXiv:2104.03016 [astro-ph.CO]}
  \BibitemShut {NoStop}%
\bibitem [{\citenamefont {Armendáriz-Picón}\ \emph
  {et~al.}(1999)\citenamefont {Armendáriz-Picón}, \citenamefont {Damour},\
  and\ \citenamefont {Mukhanov}}]{mukhanov1999}%
  \BibitemOpen
  \bibfield  {author} {\bibinfo {author} {\bibfnamefont {C.}~\bibnamefont
  {Armendáriz-Picón}}, \bibinfo {author} {\bibfnamefont {T.}~\bibnamefont
  {Damour}},\ and\ \bibinfo {author} {\bibfnamefont {V.}~\bibnamefont
  {Mukhanov}},\ }\bibfield  {title} {\bibinfo {title} {k-inflation},\ }\href
  {https://doi.org/10.1016/s0370-2693(99)00603-6} {\bibfield  {journal}
  {\bibinfo  {journal} {Physics Letters B}\ }\textbf {\bibinfo {volume}
  {458}},\ \bibinfo {pages} {209–218} (\bibinfo {year} {1999})}\BibitemShut
  {NoStop}%
\bibitem [{\citenamefont {Arnowitt}\ \emph {et~al.}(2008)\citenamefont
  {Arnowitt}, \citenamefont {Deser},\ and\ \citenamefont {Misner}}]{ADM}%
  \BibitemOpen
  \bibfield  {author} {\bibinfo {author} {\bibfnamefont {R.}~\bibnamefont
  {Arnowitt}}, \bibinfo {author} {\bibfnamefont {S.}~\bibnamefont {Deser}},\
  and\ \bibinfo {author} {\bibfnamefont {C.~W.}\ \bibnamefont {Misner}},\
  }\bibfield  {title} {\bibinfo {title} {Republication of: The dynamics of
  general relativity},\ }\href {https://doi.org/10.1007/s10714-008-0661-1}
  {\bibfield  {journal} {\bibinfo  {journal} {General Relativity and
  Gravitation}\ }\textbf {\bibinfo {volume} {40}},\ \bibinfo {pages} {1997}
  (\bibinfo {year} {2008})}\BibitemShut {NoStop}%
\bibitem [{\citenamefont {Garriga}\ and\ \citenamefont
  {Mukhanov}(1999)}]{Garriga_1999}%
  \BibitemOpen
  \bibfield  {author} {\bibinfo {author} {\bibfnamefont {J.}~\bibnamefont
  {Garriga}}\ and\ \bibinfo {author} {\bibfnamefont {V.}~\bibnamefont
  {Mukhanov}},\ }\bibfield  {title} {\bibinfo {title} {Perturbations in
  k-inflation},\ }\href {https://doi.org/10.1016/s0370-2693(99)00602-4}
  {\bibfield  {journal} {\bibinfo  {journal} {Physics Letters B}\ }\textbf
  {\bibinfo {volume} {458}},\ \bibinfo {pages} {219–225} (\bibinfo {year}
  {1999})}\BibitemShut {NoStop}%
\bibitem [{\citenamefont {Baumann}(2012)}]{Baumann2012}%
  \BibitemOpen
  \bibfield  {author} {\bibinfo {author} {\bibfnamefont {D.}~\bibnamefont
  {Baumann}},\ }\bibfield  {title} {\bibinfo {title} {{TASI Lectures on
  Inflation}},\ }\href@noop {} {\bibfield  {journal} {\bibinfo  {journal}
  {arXiv}\ } (\bibinfo {year} {2012})},\ \Eprint
  {https://arxiv.org/abs/0907.5424v2} {arXiv:0907.5424v2} \BibitemShut
  {NoStop}%
\bibitem [{\citenamefont {Handley}(2019{\natexlab{b}})}]{Handley2019}%
  \BibitemOpen
  \bibfield  {author} {\bibinfo {author} {\bibfnamefont {W.}~\bibnamefont
  {Handley}},\ }\bibfield  {title} {\bibinfo {title} {{Primordial power spectra
  for curved inflating universes}},\ }\bibfield  {journal} {\bibinfo  {journal}
  {Physics Letters D}\ }\href {https://doi.org/10.1103/PhysRevD.100.123517}
  {10.1103/PhysRevD.100.123517} (\bibinfo {year}
  {2019}{\natexlab{b}})\BibitemShut {NoStop}%
\bibitem [{\citenamefont {Cai}\ and\ \citenamefont {Xia}(2009)}]{Cai2009}%
  \BibitemOpen
  \bibfield  {author} {\bibinfo {author} {\bibfnamefont {Y.-F.}\ \bibnamefont
  {Cai}}\ and\ \bibinfo {author} {\bibfnamefont {H.-Y.}\ \bibnamefont {Xia}},\
  }\bibfield  {title} {\bibinfo {title} {{Inflation with multiple sound speeds:
  a model of multiple DBI type actions and non-Gaussianities}},\ }\href
  {https://doi.org/10.1016/j.physletb.2009.05.047} {\bibfield  {journal}
  {\bibinfo  {journal} {Physics Letters, Section B: Nuclear, Elementary
  Particle and High-Energy Physics}\ }\textbf {\bibinfo {volume} {677}},\
  \bibinfo {pages} {226} (\bibinfo {year} {2009})},\ \Eprint
  {https://arxiv.org/abs/0904.0062} {arXiv:0904.0062} \BibitemShut {NoStop}%
\bibitem [{\citenamefont {Agocs}\ \emph {et~al.}(2020)\citenamefont {Agocs},
  \citenamefont {Handley}, \citenamefont {Lasenby},\ and\ \citenamefont
  {Hobson}}]{Agocs2020}%
  \BibitemOpen
  \bibfield  {author} {\bibinfo {author} {\bibfnamefont {F.~J.}\ \bibnamefont
  {Agocs}}, \bibinfo {author} {\bibfnamefont {W.~J.}\ \bibnamefont {Handley}},
  \bibinfo {author} {\bibfnamefont {A.~N.}\ \bibnamefont {Lasenby}},\ and\
  \bibinfo {author} {\bibfnamefont {M.~P.}\ \bibnamefont {Hobson}},\ }\bibfield
   {title} {\bibinfo {title} {{Efficient method for solving highly oscillatory
  ordinary differential equations with applications to physical systems}},\
  }\href {https://doi.org/10.1103/PhysRevResearch.2.013030} {\bibfield
  {journal} {\bibinfo  {journal} {PHYSICAL REVIEW RESEARCH}\ }\textbf {\bibinfo
  {volume} {2}},\ \bibinfo {pages} {13030} (\bibinfo {year}
  {2020})}\BibitemShut {NoStop}%
\bibitem [{\citenamefont {Haddadin}\ and\ \citenamefont
  {Handley}(2021)}]{Haddadin2021}%
  \BibitemOpen
  \bibfield  {author} {\bibinfo {author} {\bibfnamefont {W.}~\bibnamefont
  {Haddadin}}\ and\ \bibinfo {author} {\bibfnamefont {W.}~\bibnamefont
  {Handley}},\ }\bibfield  {title} {\bibinfo {title} {Rapid numerical solutions
  for the mukhanov-sasaki equation},\ }\bibfield  {journal} {\bibinfo
  {journal} {Physical Review D}\ }\textbf {\bibinfo {volume} {103}},\ \href
  {https://doi.org/10.1103/physrevd.103.123513} {10.1103/physrevd.103.123513}
  (\bibinfo {year} {2021})\BibitemShut {NoStop}%
\bibitem [{\citenamefont {Martin}\ \emph {et~al.}(2013)\citenamefont {Martin},
  \citenamefont {Ringeval},\ and\ \citenamefont {Vennin}}]{Martin_2013}%
  \BibitemOpen
  \bibfield  {author} {\bibinfo {author} {\bibfnamefont {J.}~\bibnamefont
  {Martin}}, \bibinfo {author} {\bibfnamefont {C.}~\bibnamefont {Ringeval}},\
  and\ \bibinfo {author} {\bibfnamefont {V.}~\bibnamefont {Vennin}},\
  }\bibfield  {title} {\bibinfo {title} {K-inflationary power spectra at second
  order},\ }\href {https://doi.org/10.1088/1475-7516/2013/06/021} {\bibfield
  {journal} {\bibinfo  {journal} {Journal of Cosmology and Astroparticle
  Physics}\ }\textbf {\bibinfo {volume} {2013}}\bibinfo  {number} { (06)},\
  \bibinfo {pages} {021–021}}\BibitemShut {NoStop}%
\bibitem [{\citenamefont {Huang}\ and\ \citenamefont
  {Wang}(2013)}]{Huang_2013}%
  \BibitemOpen
\bibfield  {number} {  }\bibfield  {author} {\bibinfo {author} {\bibfnamefont
  {Q.-G.}\ \bibnamefont {Huang}}\ and\ \bibinfo {author} {\bibfnamefont
  {Y.}~\bibnamefont {Wang}},\ }\bibfield  {title} {\bibinfo {title} {Large
  local non-gaussianity from general ultra slow-roll inflation},\ }\href
  {https://doi.org/10.1088/1475-7516/2013/06/035} {\bibfield  {journal}
  {\bibinfo  {journal} {Journal of Cosmology and Astroparticle Physics}\
  }\textbf {\bibinfo {volume} {2013}}\bibinfo  {number} { (06)},\ \bibinfo
  {pages} {035–035}}\BibitemShut {NoStop}%
\bibitem [{\citenamefont {Handley}\ \emph {et~al.}(2014)\citenamefont
  {Handley}, \citenamefont {Brechet}, \citenamefont {Lasenby},\ and\
  \citenamefont {Hobson}}]{Handley_2014}%
  \BibitemOpen
\bibfield  {number} {  }\bibfield  {author} {\bibinfo {author} {\bibfnamefont
  {W.}~\bibnamefont {Handley}}, \bibinfo {author} {\bibfnamefont
  {S.}~\bibnamefont {Brechet}}, \bibinfo {author} {\bibfnamefont
  {A.}~\bibnamefont {Lasenby}},\ and\ \bibinfo {author} {\bibfnamefont
  {M.}~\bibnamefont {Hobson}},\ }\bibfield  {title} {\bibinfo {title} {Kinetic
  initial conditions for inflation},\ }\bibfield  {journal} {\bibinfo
  {journal} {Physical Review D}\ }\textbf {\bibinfo {volume} {89}},\ \href
  {https://doi.org/10.1103/physrevd.89.063505} {10.1103/physrevd.89.063505}
  (\bibinfo {year} {2014})\BibitemShut {NoStop}%
\bibitem [{\citenamefont {Hergt}\ \emph {et~al.}(2019)\citenamefont {Hergt},
  \citenamefont {Handley}, \citenamefont {Hobson},\ and\ \citenamefont
  {Lasenby}}]{Hergt_2019}%
  \BibitemOpen
  \bibfield  {author} {\bibinfo {author} {\bibfnamefont {L.}~\bibnamefont
  {Hergt}}, \bibinfo {author} {\bibfnamefont {W.}~\bibnamefont {Handley}},
  \bibinfo {author} {\bibfnamefont {M.}~\bibnamefont {Hobson}},\ and\ \bibinfo
  {author} {\bibfnamefont {A.}~\bibnamefont {Lasenby}},\ }\bibfield  {title}
  {\bibinfo {title} {Case for kinetically dominated initial conditions for
  inflation},\ }\bibfield  {journal} {\bibinfo  {journal} {Physical Review D}\
  }\textbf {\bibinfo {volume} {100}},\ \href
  {https://doi.org/10.1103/physrevd.100.023502} {10.1103/physrevd.100.023502}
  (\bibinfo {year} {2019})\BibitemShut {NoStop}%
\bibitem [{\citenamefont {Unnikrishnan}\ and\ \citenamefont
  {Sriramkumar}(2010)}]{Unnikrishnan2010}%
  \BibitemOpen
  \bibfield  {author} {\bibinfo {author} {\bibfnamefont {S.}~\bibnamefont
  {Unnikrishnan}}\ and\ \bibinfo {author} {\bibfnamefont {L.}~\bibnamefont
  {Sriramkumar}},\ }\bibfield  {title} {\bibinfo {title} {A note on perfect
  scalar fields},\ }\href {https://doi.org/10.1103/PhysRevD.81.103511}
  {\bibfield  {journal} {\bibinfo  {journal} {Phys. Rev. D}\ }\textbf {\bibinfo
  {volume} {81}},\ \bibinfo {pages} {103511} (\bibinfo {year}
  {2010})}\BibitemShut {NoStop}%
\bibitem [{\citenamefont {Handley}\ \emph {et~al.}(2019)\citenamefont
  {Handley}, \citenamefont {Lasenby},\ and\ \citenamefont
  {Hobson}}]{Handley_2019_logo}%
  \BibitemOpen
  \bibfield  {author} {\bibinfo {author} {\bibfnamefont {W.}~\bibnamefont
  {Handley}}, \bibinfo {author} {\bibfnamefont {A.}~\bibnamefont {Lasenby}},\
  and\ \bibinfo {author} {\bibfnamefont {M.}~\bibnamefont {Hobson}},\
  }\bibfield  {title} {\bibinfo {title} {Logolinear series expansions with
  applications to primordial cosmology},\ }\bibfield  {journal} {\bibinfo
  {journal} {Physical Review D}\ }\textbf {\bibinfo {volume} {99}},\ \href
  {https://doi.org/10.1103/physrevd.99.123512} {10.1103/physrevd.99.123512}
  (\bibinfo {year} {2019})\BibitemShut {NoStop}%
\bibitem [{\citenamefont {Blas}\ \emph {et~al.}(2011)\citenamefont {Blas},
  \citenamefont {Lesgourgues},\ and\ \citenamefont {Tram}}]{2011class}%
  \BibitemOpen
  \bibfield  {author} {\bibinfo {author} {\bibfnamefont {D.}~\bibnamefont
  {Blas}}, \bibinfo {author} {\bibfnamefont {J.}~\bibnamefont {Lesgourgues}},\
  and\ \bibinfo {author} {\bibfnamefont {T.}~\bibnamefont {Tram}},\ }\bibfield
  {title} {\bibinfo {title} {The cosmic linear anisotropy solving system
  (class). part ii: Approximation schemes},\ }\href
  {https://doi.org/10.1088/1475-7516/2011/07/034} {\bibfield  {journal}
  {\bibinfo  {journal} {Journal of Cosmology and Astroparticle Physics}\
  }\textbf {\bibinfo {volume} {2011}}\bibinfo  {number} { (07)},\ \bibinfo
  {pages} {034–034}}\BibitemShut {NoStop}%
\bibitem [{\citenamefont
  {Handley}(2019{\natexlab{c}})}]{Handley2019anesthetic}%
  \BibitemOpen
\bibfield  {number} {  }\bibfield  {author} {\bibinfo {author} {\bibfnamefont
  {W.}~\bibnamefont {Handley}},\ }\bibfield  {title} {\bibinfo {title}
  {anesthetic: nested sampling visualisation},\ }\href
  {https://doi.org/10.21105/joss.01414} {\bibfield  {journal} {\bibinfo
  {journal} {Journal of Open Source Software}\ }\textbf {\bibinfo {volume}
  {4}},\ \bibinfo {pages} {1414} (\bibinfo {year}
  {2019}{\natexlab{c}})}\BibitemShut {NoStop}%
\end{thebibliography}%

\end{document}